# Direct transfer of classical non-separable state into hybrid entangled two photon state

M. V. JABIR,[1*] N. APURV CHAITANYA,[1] MANOJ MATHEW,[2] G. K. SAMANTA[1]

[1]*Photonic Sciences Lab., Physical Research Laboratory, Navarangpura, Ahmedabad 380009, Gujarat, India.*
[2] *National Centre for Biological Sciences, Bengaluru, India.*

*Corresponding author: jabir@prl.res.in*

**Hybrid entangled states, having entanglement between different degrees-of-freedom (DoF) of a particle pair, are of great interest for quantum information science and communication protocols. Among different DoFs, the hybrid entangled states encoded with polarization and orbital angular momentum (OAM) allow the generation of qubit-qudit entangled states, macroscopic entanglement with very high quanta of OAM and improvement in angular resolution in remote sensing. Till date, such hybrid entangled states are generated by using a high-fidelity polarization entangled state and subsequent imprinting of chosen amount of OAM using suitable mode converters such as spatial light modulator in complicated experimental schemes. Given that the entangled sources have feeble number of photons, loss of photons during imprinting of OAM using diffractive optical elements limits the use of such hybrid state for practical applications. Here we report, on a simple experimental scheme to generate hybrid entangled state in polarization and OAM through direct transfer of classical non-separable state of the pump beam in parametric down conversion process. As a proof of principle, using local non-separable pump state of OAM mode $l$=3, we have produced quantum hybrid entangled state with entanglement witness parameter of $\widehat{W}$= 1.25±0.03 violating by 8 standard deviation. The generic scheme can be used to produce hybrid entangled state between two photons differing by any quantum number through proper choice of non-separable state of the pump beam.**

## 1. INTRODUCTION

Entanglement, the quintessential strong non-classical correlations in joint measurement of at least two separate quantum systems, plays a critical role in many important applications in quantum information processing, including quantum communication [1], quantum computation [2], quantum cryptography [3], dense coding [4] and teleportation [5]. Typically, in photonic quantum optics, spontaneous parametric down-conversion (SPDC) is used to produce correlated photon pairs [6-9] with many accessible degree-of-freedom (DoF) that can be exploited for the production of entanglement. With the first demonstration of entanglement with polarization DoF [10,11], recent advancement in quantum optics have provided intrinsic entanglement (entanglement in variety of DoFs such as orbital angular momentum (OAM) [12], energy time [13], time bin [14], and many more[15]), hyperentanglement [16] (entanglement in every DoFs) and hybrid entanglement (entanglement between different DoFs of a pair of particles). While these entangled states have various applications, the hybrid entangled states encoded with polarization and OAM, in particular, allow the generation of qubit-qudit entangled states [17] for quantum information, macroscopic entanglement with very high quanta of OAM [18], important for quantum information science, and supersensitive measurement of angular displacement in remote sensing [19].

Typically, hybrid entangled states encoded with polarization and OAM are generated through the imprinting of chosen amount of OAM to a high-fidelity polarization entangled state using mode converters, such as, spatial light modulator (SLM) [20] and q-plate [21], in complicated experimental schemes. As compared to all mode converters, the SLM have many advantages in terms of dynamic variation in OAM, accessibility to very high OAM and flexibility in imprinting two particles with an arbitrarily high difference in OAM [18]. However, the diffraction based OAM imprinting process of the SLM reduces the overall number of photons and thus limiting the use of hybrid state for practical applications requiring entangled state with high brightness. To circumvent such problem, as such, it is imperative to device alternative techniques to produce hybrid entangled states in simple experimental scheme.

Entanglement properties of the paired photons generated through SPDC process are highly influenced by different crystal parameters including birefringence and length, and the spatial structure of the pump beam [22, 23]. Recent studies have shown that the transfer of pump properties such as non-diffraction [24], intensity distribution and phase structure [25, 26] into the transverse amplitude of heralded single photons. Therefore, one can in principle manipulate the pump beam to directly generate hybrid entangle state through SPDC process.

On the other hand, light beams with non-separable states in polarization and OAM [27, 28] have attracted a great deal of interest due to its violation of Bell like inequality [29, 30]. Here we propose, for the first time to the best of our knowledge, direct transfer of non-separable laser beams into hybrid entangle two photon state in a simple experimental scheme. As a proof of principle, pumping the contiguous nonlinear crystals with classical non-separable pump beam of OAM mode $l$=1 and 3, and characterizing the quantum state through violation of Bell's inequalities and the measurement of entanglement witness operator (W) for twin photons we showed that the generated two photons are entangled in both polarization and OAM. The concept is generic and can be used for hybrid entanglement with higher OAM, and photons with arbitrarily high difference in OAM through proper choice of non-separable state of the pump beam. The concern of rapidly decreasing efficiency of the down conversion process for the direct generation of entanglement of higher OAM, can be overcome by OAM independent beam size of the non-separable state using the scheme used in Ref. [23, 31].

## 2. EXPERIMENTAL SETUP

The schematic of the experimental setup is shown in Fig. 1(a). A continuous-wave, single-frequency (<12 MHz) UV laser providing 70 mW of output power at 405 nm in Gaussian spatial profile is used as a pump laser. The laser power to the experiment is controlled using a half-wave plate ($\lambda/2$) and a polarizing beam splitter (PBS1) cube. A second $\lambda/2$ plate placed after PBS1 converts the linearly polarized Gaussian beam represented by the state, $|H\ 0\rangle$ (here, the first and second terms of the ket represent polarization and OAM of the beam respectively) in to $\psi 1 = 1/\sqrt{2}(|H\ 0\rangle + |V\ 0\rangle)$. Here, $H$ and $V$ represents horizontal and vertical polarization respectively and the Gaussian beam has OAM mode of, $l$=0. To prepare classical non-separable state, the pump beam is passed through a polarization Signac interferometer comprising with PBS2, three mirrors (M) and a polarization independent spiral phase plate, SPP. The PBS2 splits the pump state, $\psi 1$ in two counter propagating beams in the Sagnac interferometer with $|H\ 0\rangle$ and $|V\ 0\rangle$ beams propagating in counter clock-wise (CCW) and clock-wise (CW) directions respectively. After a round trip both the beams recombine in the PBS2 and produce output state same as that of the input state, $\psi 1$. However, due to the presence of SPP that converts Gaussian beam ($l$=0) into optical vortex of OAM mode, ±$l$, the output state of the Sagnac interferometer will be different than that of the input state. Depending on the direction of thickness variation of the SPP, if the CCW beam, $|H\ 0\rangle$, while passing through the SPP in the Sagnac interferometer acquires spiral phase corresponding to an optical vortex of order +$l$ (say) then the CW beam, $|V\ 0\rangle$, will acquire optical vortex of order -$l$ and vice versa. As a result, the output of the Sagnac interferometer can be represented by the classical non-separable state, $\psi 2 = 1/\sqrt{2}(|H\ l\rangle + e^{-i\phi}|V\ -l\rangle)$. The phase factor, $\phi$, arises due to the asymmetric positioning of the SPP inside the Sagnac interferometer. Two contiguous BIBO (bismuth borate) crystals each of 0.6-mm thick and 10 x 10 mm² in aperture with optic axes aligned in perpendicular planes, is used as nonlinear crystal for SPDC process. Both the crystals are cut with, $\theta$=151.7° ($\phi$=90°) in optical $yz$-plane for perfect phase-matching of non-collinear type-I ($o \rightarrow e+e$) degenerate down converted photons at 810 nm in a cone of half-opening angle~3° for normal incidence of pump. Orthogonal positioning of the optic axes of the BIBO crystals facilitate the pump photons in both $|H\rangle$ and $|V\rangle$ polarization states to produce respective non-collinear SPDC photons in concentric cones around the direction of the pump beam. For entanglement studies, we select two diametrically opposite points of the SPDC ring [red circle in Fig. 1(a)] in the horizontal plane and named them as Arm-1 and Arm-2. In Arm-1, we conditioned one of the down-converted photons (say idler photon) using a hard aperture of diameter ~460 µm and multimode fiber [25].

To herald its partner photon (here signal) we collected the photons in Arm-2 using a collimator with an opening diameter of ~460 µm and a multi-mode fiber assembly placed on $x$-$y$ scanning stage. The photons of each arm are analyzed through coincidence count using a combination of single photon counting module (SPCM) and a time-to-digital converter (TDC). A time window of 2.5 ns was used to measure the coincidence counts. Quarter-wave plate, $\lambda/4$, and analyzers, A, [comprised with a $\lambda/2$ plate and PBS as shown in Fig. 1(b)] are used to analyze the photons in polarization basis. The interference filter (IF) with transmission bandwidth of ~10 nm with central wavelength at 810 nm is used to extracts degenerate photons from broad spectrum of SPDC process. Figure 1(c)-(e) show the spatial distribution of the non-separable state of pump beam, conditioned idler and heralded signal photons respectively.

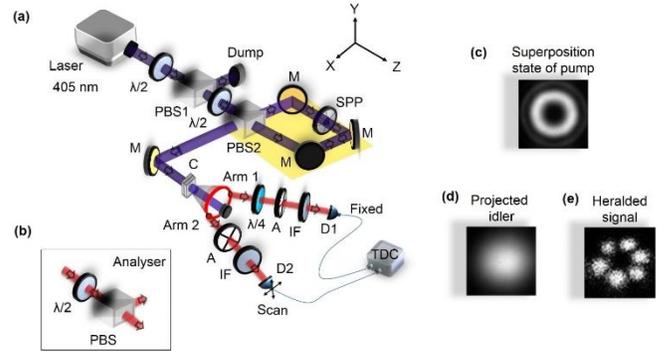

Fig. 1. **Direct generation of hybrid entangled two photon state**. (a) Schematic of the experimental setup. Laser 405 nm, continuous wave single frequency diode laser at 405 nm providing 70 mW of output power; $\lambda/2$, half-wave plate; PBS1-2, polarizing beam splitter cube; SPP, spiral phase plates; M, mirrors; schematic marked in yellow represents polarization Signac interferometer; C, dual-BIBO crystal having optic axis orthogonal to each other for the generation of entangled photons; A, analyser; $\lambda/4$, quarter wave plate; IF, interference filter; D1-2, single photon counting module (SPCM); TDC, time-to-digital converter. (b) Analyser comprises with PBS and $\lambda/2$. Spatial distribution of the (c) non-separable pump beam, (d) conditioned idler photon and (e) heralded signal photons.

## 3. RESULTS AND DISCUSSIONS

In non-collinear SPDC process, where high energy photon owing to energy conservation splits into two low energy photons known as signal and idler, the generated entangled photons are distributed in a ring with signal and idler photons laying in diametrically opposite points [see Fig. 1(a)]. To study the entanglement quality of the down converted photons in the current experimental scheme we pumped the dual-BIBO crystal with input state $\psi 1 = 1/\sqrt{2}(|H\ 0\rangle + |V\ 0\rangle)$ in Gaussian spatial intensity distribution by removing the SPP from the Sagnac interferometer. Due to orthogonal positioning of the optic axes of the crystals, if the first crystal produces down converted photons of the pump photons of $|H\rangle$ polarization state into $|VV\rangle$ owing to type-I phase-matching, the second crystal converts pump photons of $|V\rangle$ polarization state into down converted photons in $|HH\rangle$ state and vice versa. Since the crystal thickness is small and the coherence length of the pump laser is very high (~25 m) the photons generated from the both crystals are highly indistinguishable in space and time. Therefore, one can write the output state, $\psi$ of the down converted photons as superposition of individual states $|HH\rangle$ and $|VV\rangle$. However, one need to determine the state and the quality of polarization entanglement in two photon state. In doing so we used standard coincidence measurement technique and recorded the two-photon interference in terms of photon coincidence between the twin photons distributed in Arm-1 and Arm-2 under two non-orthogonal projection bases, H/V (horizontal/vertical) and D/A (diagonal/anti-diagonal) using two polarization analyzers as the quantum state analyzer with the results shown in Fig. 2(a). As expected, the normalized coincidence rate measured 10 sec show the expected sinusoidal variation with angle of the quantum state analyzer with fringe visibilities 99.7±0.03% and 96.9±0.04% for H (red dots) and D (black dots) bases respectively. The measured visibilities under both basis are higher than 71% [32], large enough to violate Bell's inequality. However, using the coincidence rates we measured the Bell's parameter to be S= 2.73±0.04 indicating the polarization entanglement of the generated two-photon state. We also constructed the density matrix of the state using linear tomographic technique [33]. Figure 2(b) shows the graphical representation of density matrix of the generated state. From this analysis we determine the state to be $\psi 3 = 1/\sqrt{2}(|HH\rangle - |VV\rangle)$ and fidelity is 0.992.

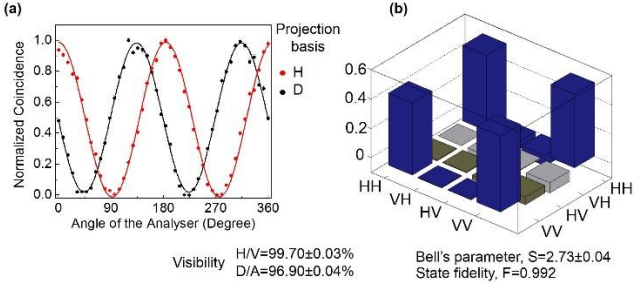

Fig. 2. **Study of two-photon polarization entanglement and identification of the state**. (a) Visibility graph for polarization entangled state at H (red dots) and D (black dots) bases. (b) Graphical representation of the density matrix obtained from linear tomographic technique for polarization entangled state.

For hybrid-entanglement, we prepared the classical pump beam in non-separable state [28] in OAM and polarization DoF by placing SPP inside the polarization Sagnac interferometer with phase variation corresponding to OAM mode of $l$=1, 2 and 3. The non-separable state of the pump beam incident to the nonlinear crystal can be expressed, $\psi 2 = 1/\sqrt{2}(|H\ l\rangle + e^{-i\phi}|V\ -l\rangle)$ with intensity distribution shown in Fig. 1(c). To verify the non-separability, measurement in one DoF influence the outcome of the measurement in other DoF, we projected the pump state having OAM mode of $l$=1 and 3 at different polarization states, horizontal ($|H\rangle$), vertical ($|V\rangle$), anti-diagonal, ($|A\rangle$), diagonal ($|D\rangle$), left circular ($|L\rangle$) and right circular ($|R\rangle$) in Poincaré sphere and recorded the intensity of the beam with the results shown in Fig. 3. As evident from Fig. 3, the projection of the pump state on $|H\rangle$ and $|V\rangle$ states result in vortex intensity profile with OAM order $l$ and $-l$ respectively. However, the projection on $|A\rangle$, $|D\rangle$, $|L\rangle$ and $|R\rangle$ states produce intensity distribution corresponding to the superposition of two opposite helical wavefronts of OAM order $l$ resulting a ring lattice structure containing $2l$ number of petals at different orientation. All these projected intensity distribution can be represented by different points on LG-Bloch (Poincaré) sphere. The change in the images of Fig. 3(a) and Fig. 3(b) representing the projection of the pump state corresponding to OAM mode of $l$=1 and 3 respectively corresponding to the change in polarization projection, verifies the generation of non-separable states. The inset image of Fig. 3(a) and 3(b) show intensity profile of the pump state without any projection.

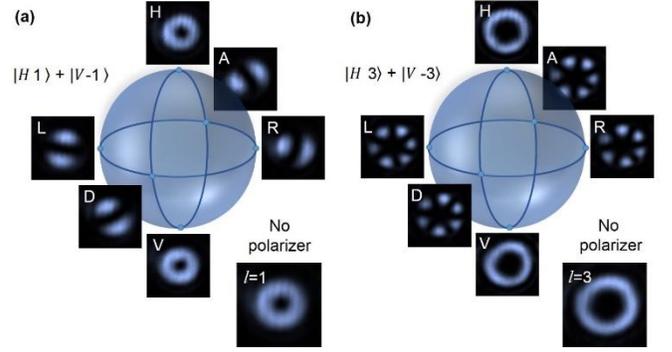

Fig. 3. **Intensity profiles of the non-separable states of the pump beam for two different OAM modes**. Depending on the projection of the beam in different polarization states, $|H\rangle$, $|A\rangle$, $|R\rangle$, $|V\rangle$, $|D\rangle$ and $|L\rangle$ (as shown by the white letters on the images) on the Poincaré sphere, the mode pattern of the beam recorded by the CCD camera change to different points on LG-Bloch sphere for (**a**) $1/\sqrt{2}(|H\ 1\rangle + e^{-i\phi}|V\ -1\rangle)$ and (**b**) $1/\sqrt{2}(|H\ 3\rangle + e^{-i\phi}|V\ -3\rangle)$. Inset images are intensity distribution of the non-separable states for OAM mode $l$=1 and $l$=3 without any polarization projection.

With successful generation and verification of the non-separable state, we pumped the dual-BIBO crystal with the state, $\psi 2 = 1/\sqrt{2}(|H\ l\rangle + e^{-i\phi}|V-l\rangle)$, for direct transfer of classical non-separable state into hybrid entangled two photon states. According to the OAM conservation in nonlinear processes [34], the OAM of the pump photon should be equal to the sum of the OAMs of the generated signal and idler photons, $l_p = l_s + l_i$ [35, 36]. Here, $l_p$, $l_s$ and $l_i$ are the OAM of pump, signal and idler photons respectively. Since the OAM conservation law does not put any selection rule for the OAMs of the signal and idler photons, the OAMs of signal and idler can have arbitrary values with random variation owing to the conservation law. However, if we force signal or idler to carry fixed OAM value, then its partner photon will have a particular OAM with certainty. For example, if we project either signal or idler photons into Gaussian mode ($l$=0), then the OAM of the idler or signal photon will be equal to that of pump photon indicating the possibility of direct transfer of the non-separable state in OAM and polarization DoF of the pump into one of the down converted photon. Therefore, in the present experiment, the state of the paired photons (considering idler photons in Gaussian mode) generated from the dual-BIBO crystal while pumped with non-separable state, $\psi 2 = 1/\sqrt{2}(|H\ l\rangle + e^{-i\phi}|V-l\rangle)$ can be written as, $\psi 4 = \alpha|V\ l\rangle + e^{-i\phi}\beta|H\ -l\rangle)$ while the signal photon is projected by the analyzer in the D polarization. However, if the idler photon is projected in D polarization (for example) the paired photon state will transformed in to $\psi 5 = |D\rangle \otimes [\alpha|l\rangle + \beta e^{-i\phi}|-l\rangle]$, where, $\alpha$, $\beta$ and $\phi$ are real constants and $\alpha^2 + \beta^2 = 1$. As a proof of principle, in the present experimental scheme, we pumped the crystal with non-separable state, $\psi 2$ for two different values of OAM ($l$=1, 3) as shown in Fig. 3, and projected the photon (idler) of Arm-1 into Gaussian mode ($l$=0) using a lens and multi-mode fiber similar to Ref. [24, 25] and measured the spatial distribution of the heralded photon (signal) in the form of coincidence counts per 10s in the transverse [x-y, see Fig 1(a)] plane of Arm-2 with the results shown in Fig. 4. As evident from the gallery of images of Fig. 4(a) and Fig. 4(b) representing the spatial distribution of the heralded signal photon for the non-separable state corresponding to OAM mode of $l$=1 and $l$=3 respectively, the projection of the idler photon in Arm-1 to different states ($|H\rangle$, $|A\rangle$, $|R\rangle$, $|V\rangle$, $|D\rangle$ and $|L\rangle$ (as marked by the white letters in the images) in the polarization Poincaré sphere with the help of λ/4 plate and analyzer, A, directly projects the probability distribution of the heralded signal photons in Arm-2 in a ring lattice spatial structure with $2l$ number of petals to different points on the LG-Bloch sphere. Such observation intuitively gives the impression of generation of hybrid entangled two photon state through the direct transfer of non-separable state of the pump.

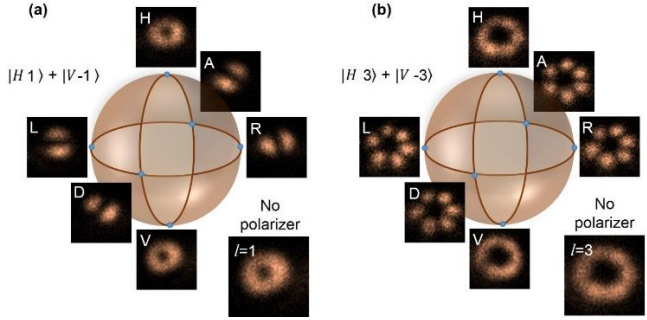

Fig. 4. **Gallery of images representing the probability distribution of the heralded signal photons**. Depending on the polarization of the idler photon projected (white letters in the images) at different points in the polarization Poincaré sphere, the spatial distribution of the heralded signal change into different mode patterns in the LG-Bloch sphere. The sequence of coincidence images of the heralded signal due to particular polarization of its partner (idler) photon while pumping with non-separable state of (a) first order, and (b) third order LG modes around the meridian (vertical circle) and the equator (horizontal circle) in the LG-Bloch sphere confirms the generation of hybrid entangled two photon state encoded in polarization and OAM. In absence of any polarizer in the path of the idler photon, the spatial distribution of the heralded signal photon (inset) shows a statistical mixture of all states of the LG-Bloch sphere.

However, for confirmation and quantitative study of the entanglement we explore the features of the LG modes. It is well known that the superposition of any two equal OAM modes with opposite helicities, $|LG_{\pm l}\rangle = |l\rangle + e^{-i\phi}|-l\rangle$ results in radially symmetric ring lattice with $2l$ number of petals. However, the relative phase, $\phi$ between the two OAM modes results in spatial rotation of $\frac{\phi}{2\pi}\frac{360^o}{2l}$ [5] which can in principle be used to identify and discriminate between different superpositions of the OAM modes. To distinguish the spatial rotation and therefore different superpositions for the verification of entanglement in OAM and polarization DoF we have evaluated the coincidence of the heralded signal photons per angular region, $\theta$, from the coincidence images (see insets of Fig. 5) for the idler polarization in mutually unbiased bases, $|A\rangle, |D\rangle$ and $|R, |L\rangle$ for the pump OAM mode, $l=1$ and $l=3$ with the results shown in Fig. 5(a) and Fig. 5(b) respectively. Let's assume that at an angle $\theta$, we have the maximum coincidence for anti-diagonal (A) projection. The $n^{th}$ maxima in the same projection appears at an angle, $\theta + n\frac{360^o}{2l}$, where $n$ is an integer in the range, $1 \leq n \leq 2l$. The angular separation between two consecutive maxima is, $\frac{360^o}{2l}$. On the other hand, in mutually unbiased (say, L) and orthogonal (D) basis the maxima will shift to an angle $\theta + \frac{45^o}{l}$ and $\theta + \frac{90^o}{l}$ respectively. As evident from Fig. 5(a), for OAM mode, $l=1$, two consecutive maxima in A-projection (black dots) occur at $\theta=0^o$ and $\theta + \frac{360^o}{2l}=180^o$ and the maxima in the L- (blue dots) and D- (red dots) projections have angular shift of angle $\theta + \frac{45^o}{l}=45^o$ and $\theta + \frac{90^o}{l}=90^o$ respectively with respect to A-projection. Similarly in case $l=3$, as evident from Fig. 5(b), we observe two consecutive maxima in same projection basis to have an angular separation of $60^o$ and the maxima in the L- (blue dots) and D- (red dots) projections have angular shift of angle $15^o$ and $30^o$ respectively with respect to maxima in A-projection. Such spatial rotation of the OAM mode of the heralded signal photon for the projection of the idler at different points in the polarization Poincaré sphere confirms the entanglement in both OAM and polarization DoF. However, to estimate the entanglement quality we have calculated the entanglement witness operator, $\widehat{W} = V_{R/L} + V_{D/A}$ using the quantum interference visibilities, $V_{R/L}$ and $V_{D/A}$ in two mutually unbiased bases R/L and D/A respectively. For all separable states, the entanglement witness operator satisfies the inequality, $\widehat{W} \leq 1$ and exceeding the limit verifies entanglement [20, 37]. Using the values of $V_{R/L}$ and $V_{D/A}$ we estimate the entanglement witness operator for OAM mode $l=1, 2,$ and $l=3$ to have a value $\widehat{W}= 1.56\pm0.04, 1.40\pm0.04$ and $1.25\pm0.03$, clearly violating the inequality by more than 14, 10 and 8 standard deviation respectively. It is to be noted that we did not apply any background correction to the experimental results. Slight lower violation of inequality in case $l=3$ with respect to that of $l=1$ can be attributed to error in visibility data due low signal to noise ratio in the spatial distribution of the lower number of down converted photons. The stronger violation requires increase in the number of down converted photons. However, the present study confirms generated twin photons are entangled in both OAM and polarization DoF.

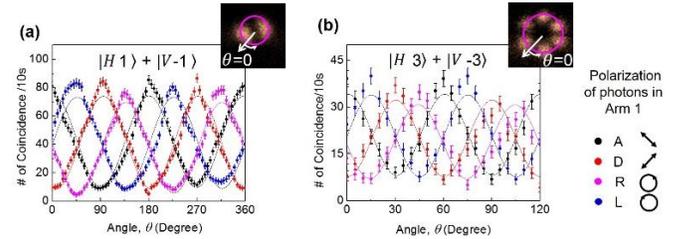

Fig. 5. **Distribution of heralded signal photon number per angular region**. Coincidence per angular region for A (black dots), D (red dots), R (pink dots) and L (blue dots) projections of the idler photon for OAM (a) $l=1$, (b) for $l=3$. The lines are theoretical fit to experimental data. (Inset) Image of the probability distribution of the heralded signal photons. Errors are estimated from iteration.

## 4. CONCLUSIONS

In conclusion, we have successfully demonstrate a novel scheme of generating hybrid entangled two photon state. Pumping a contiguous nonlinear crystals using non-separable state of pump beam of OAM mode of $l=1$ and 3 we have generated two photon hybrid entangled state. Characterization of the generated quantum state through tomography, and violation of Bell's inequality parameter shows high quality polarization entanglement whereas, the measurement of entangled witness parameter verifies the generation of two photons state entangled in both OAM and polarization DoFs. The concept is generic and can be used to produce hybrid entangled state with higher quanta of OAM and photons with arbitrarily high difference in OAM through the proper selection of the nonclassical state of the pump beam.